\documentclass{emulateapj}

\pdfoutput=1
\usepackage{graphicx}
\usepackage{natbib}
\usepackage{amsmath}
\usepackage{amssymb}

\begin{document}


\bibliographystyle{apj}

\shortauthors{Chakrabarti \& Blitz}  \righthead{Tidal Imprints Of A Dark Sub-Halo II.}

\title{Tidal Imprints Of A Dark Sub-Halo On The Outskirts Of The Milky Way. II. Perturber Azimuth}

\author{ Sukanya Chakrabarti\altaffilmark{1,2} \& Leo Blitz \altaffilmark{1}} \altaffiltext{1}{Astronomy Department, UC Berkeley, 601 Campbell Hall, Berkeley, CA 94720 USA; sukanya@astro.berkeley.edu}

\altaffiltext{2} {UC Presidential Postdoctoral Fellow}

\slugcomment{Submitted to ApJ, June 24, 2010}


\begin{abstract}
We extend our analysis of the observed disturbances on the outskirts of the HI disk of the Milky Way.  We employ the additional constraints of the phase of the modes of the observed HI image and asymmetry in the radial velocity field to derive the azimuth of the perturber inferred to be responsible for the disturbances in the HI disk.  Specifically, we carry out a modal analysis of the phase of the disturbances in the HI image and in high resolution Smoothed Particle Hydrodynamics (SPH) simulations of a Milky Way-like galaxy tidally interacting with dark perturbers, the relative offset of which we utilize to derive the perturber azimuth.  Under the assumption that the asymmetry in the radial velocity is due to the perturber, we derive the best-fit to the radial velocity at $l=0,180$ and use this constraint to also derive the perturber azimuth.  To make a direct connection with observations, we express our results in sun-centered coordinates, predicting that the perturber responsible for the observed disturbances is between $-50 \la l \la -10$.  We show explicitly that the phase of the disturbances in the outskirts of simulated galaxies at the time that best fits the Fourier amplitudes, our primary metric for the azimuth determination, is relatively insensitive to the equation of state (for the range of gas fractions of local spirals).  Our calculations here represent our continuing efforts to develop the ``Tidal Analysis'' method of Chakrabarti \& Blitz (2009; CB09).  CB09 employed SPH simulations to examine tidal interactions between perturbing dark sub-halos and the Milky Way.  They found that the amplitudes of the Fourier modes of the observed planar disturbances are best-fit by a perturbing dark sub-halo with mass $\sim$ one-hundredth that of the Milky Way, and a pericentric approach distance of $\sim 5-10~\rm kpc$.  The overarching goal of this work is to attempt to outline an alternate procedure to optical studies for characterizing and potentially discovering dwarf galaxies -- whereby one can approximately infer the azimuthal location of a perturber, its mass and pericentric distance (CB09) from analysis of its tidal gravitational imprints on the HI disk of the primary galaxy. 

\end{abstract}


\section{Introduction}

The detection and characterization of the ubiquitous dark matter in the universe is an outstanding problem in modern astrophysics.  The overabundance of dark sub-halos (or dim dwarf galaxies) in theoretical simulations of the Milky Way (Klypin et al. 1999; Moore et al. 1999; Kravtsov et al. 2004) relative to observations of Local Group dwarfs has come to be known as the missing satellites problem.  Recent discoveries of dark matter dominated Milky Way satellites (Willman et al. 2005; Belokurov et al. 2006; Walsh et al. 2007) have prompted much theoretical work on dwarf galaxies, mainly focused on the formation of dwarf galaxies (Governato et al. 2010, and references therein).  These recent works have addressed a number of important problems, including the formation of bulgeless dwarfs (e.g. Governato et al. 2010).  Nonetheless, the so-called missing satellites problem is still wanting a solution, and an approach that focuses on the dynamical interactions between dwarf galaxies and the primary galaxy would be extremely valuable towards this end.  

The excess of sub-structure in simulations relative to observations of dwarf galaxies prompted us to ask -- can dark dwarf galaxies be characterized from disturbances, left by tidal imprints of their passage, in the cold gas on the outskirts of galaxies?  Thus, we were motivated to analyze observed perturbations (Levine, Blitz \& Heiles 2006a) on the outskirts of the gas disk of the Milky Way in an earlier paper (Chakrabarti \& Blitz 2009; henceforth CB09), with the hypothesis that these disturbances arise from an external perturber that tidally interacted with the Milky Way.  By quantitative comparison with a large set of high-resolution SPH simulations of a Milky Way-like galaxy tidally interacting with dark sub-halos, CB09 found that the best-fit to the Fourier amplitudes of the planar disturbances occurred for sub-halos with mass ratio one-hundredth that of the Milky Way and a distance of closest approach $\sim 5-10~\rm kpc$.  A fundamental recognition of CB09 was that the nonlinear response of the primary galaxy along the orbit of the satellite allows one to break the degeneracy between the mass and distance cubed in the tidal force -- yielding a dynamical measure of the masses of satellites from analysis of the Fourier modes of the gas surface density.  In CB09, we did not calculate the azimuthal location of this perturber.  In this paper, we further develop the ``Tidal Analysis'' method of CB09 to outline our inference of the azimuthal location of perturbers from their tidal effects on gas disks.  Specifically, we employ the additional constraints of the phase variation of the modes of the surface density in both the simulations and in the observations, as well as asymmetries in the radial velocity field, to derive the perturber azimuth.  To make a direct connnection with observations, we express our results in sun-centered coordinates.

The goal of this work is to attempt to delineate a procedure by which one characterizes dark perturbers, in particular to infer the azimuthal location of perturbers from analysis of the observed disturbances on the gas disks of galaxies.  If successful, this method may allow for dark (or effectively dark, i.e., with large mass-to-light ratios) sub-halos to be discovered in infrared surveys, even if they are not very deep.  In addition to optical studies, cold dark matter (CDM) sub-structure in galaxies can alternately be studied via gravitational lensing (Dalal \& Kochanek 2002), a method that like the tidal analysis method of CB09 does not rely on the stellar light distribution, but is subject to uncertainties in the projected mass distribution.  With the advent of upcoming radio surveys by SKA and ALMA, we wish to further develop our alternate approach of characterizing CDM sub-structure by analysis of observed perturbations in the gas surface density, as the new generation of radio telescopes will be able to image extended HI disks out to higher redshift, promising an independent method of understanding the redshift evolution of CDM sub-structure.  Our hypothesis here is a simple one, namely that the disturbances are due to an external perturber.  We examine this assumption in detail in a future work where we examine the effects of multiple perturbers and asymmetries in the dark matter halo.  This paper is organized as follows:  in \S 2, we present our results and analysis with a brief synopsis of methods used; in \S 3 we discuss our results, future work and caveats, and we conclude in \S 4.

\section{Results}

The simulations and methodology have been previously presented in CB09.  We briefly review them here.  We used the GADGET-2 simulation code (Springel 2005) to examine tidal interactions between perturbers and the Milky Way disk in CB09 for a set of $\sim$ 50 simulations to find that the amplitudes of the the low-order ($m=1-4$) Fourier modes of the HI data are best-fit by a 1:100 perturber with a pericentric approach distance of $R_{\rm peri} \sim 5~\rm kpc$, at a current radial location $\sim 80~\rm kpc$.  By plotting the variances of the $m=1$ mode relative to the variance of modes $1-4$ (a variance vs. variance plot necessarily situates the best-fit simulations close to the origin), we showed that variations in initial conditions of the simulated Milky Way disk, such as gas fraction and equation of state, do not produce effects that are statistically significant.  (This is true as long as the gas fraction is varied to within the range of typical local spirals)  We showed explicitly that varying orbital inclination for parabolic orbits led to small changes on the variance vs variance plot.  A corollary of this finding is that the orbital history cannot be uniquely determined due to the (near) degeneracy in resultant Fourier amplitudes of the gas disturbances.  Significant changes on this variance vs variance plot were primarily driven by changes in gross parameters, such the mass of the perturber and distance of closest approach.  

We also showed that it is possible to break the degeneracy between the mass and distance cubed in the tidal force under certain conditions.  This can be done if the time it takes the satellite to traverse the distance from $R_{0}(M_{s})$ (the minimum distance at which a satellite of mass $M_{s}$ starts to raise tides on the primary galaxy) to the pericentric approach distance $R_{\rm peri}$ is greater than the time it takes for the gas to shock.  If so, then disturbances will first form on the scale of $R_{0}(M_{s})$ (at which point the pull of the satellite starts to overwhelm the local gravity in the primary galaxy), resulting in a visibly different response for two satellites of differing mass that exert an $\it{equivalent}$ tidal force at their respective pericentric approach distances (Figure 3 in CB09).  This is therefore a dynamical method of determining the masses of satellites, from analysis of the observed HI map, which does not require knowledge of the stellar velocity dispersion (see Wolf et al. 2009 for a review of masses of Milky Way satellites from the observed stellar velocity dispersion).  

These two things taken together, i.e., the possibility of the determination of the mass of the satellite and the pericentric distance when one is not in the impulse approximation, and the relative lack of sensitivity on the initial conditions of the primary galaxy and orbital inclination -- are what allow us to characterize dark perturbers from their tidal imprints on gas disks.  In a forthcoming paper, we perform a linear analysis and show that the $M_{s}$, $R_{\rm peri}$ degeneracy can indeed be broken for any orbit.  

Our calculations track the locations of the particles used to simulate the Milky Way, as well as the perturber.  Therefore, we were able to determine the current radial location of the perturber at the best fit time snapshots in CB09.  The notion of a best-fit time here derives from the fact that the disturbances in the gaseous component damp out on the order of a dynamical time.  Our metric of comparison to the data were the azimuthally integrated radially varying Fourier modes -- which yielded the mass of the perturber and distance of closest approach.  

We did not make use of all the information present in the images of the data and simulations in our analysis in CB09 -- an image is composed of the Fourier amplitudes as well as the phases of the modes.  Since the simulations are performed in the center-of-mass frame, we could not immediately determine the azimuthal location of the perturber in sun-centered (or galactocentric) coordinates.  Our method for determining the azimuthal location of the perturber is as follows:  we compute the phase of individual modes of the simulations and the data for the sub-set of simulations (and at the best-fit times for a given simulation) that give the best-fit to the amplitudes of the Fourier modes as determined by CB09.  Choosing this sub-set guarantees that we consider those class of simulations with perturber masses and pericentric approach distances that reasonably fit the amplitudes of the observed disturbances.  For projected gas surface density denoted $\Sigma(r,\phi)$, we calculate the phase of individual modes ``m'' by taking the FFT:
\begin{equation}
\phi(r,m) = \arctan\frac{\left[-Imag~FFT~\Sigma(r,\phi) \right]}{\left[-Re~FFT~\Sigma(r,\phi)\right]} \; .
\end{equation}
The phase of the modes contains information on the shape of the spiral planform.  Tightly wrapped spirals produced by self-excited spiral structure inside of the Inner and Outer Lindblad Resonances will have a sharp gradient in the phase, while open spirals as produced by tidal interactions will have a flatter profile (Shu 1984).  This type of Fourier analysis is similar to observational analyses by Considere \& Athanassoula (1982) who fit logarithmic spirals to study the distribution of HII regions in images of external galaxies (for more detailed recent examples see Garcia-Gomez et al. 2004; Block et al. 2009).  Since the simulations are not performed in the observational frame, there is no reason why -- even if the phase variation of the simulations ($\phi^{\rm sim}(r,m)$) and data ($\phi^{\rm data}(r,m)$) are similar in shape -- that the relative offset between $\phi^{\rm sim}(r,m)$ and $\phi^{\rm data}(r,m)$ will be zero.  We make use of this relative offset between $\phi^{\rm sim}(r,m)$ and $\phi^{\rm data}(r,m)$ to translate our determination of the perturber azimuth in the simulation frame to the observational frame.  This procedure is similar to visual matching of (dominant) features in the data and simulations.  We then use standard conversions to express our results in the sun-centered frame.

\begin{figure}
\begin{center}
\includegraphics[scale=0.6]{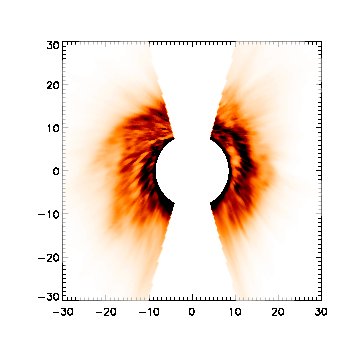}
\includegraphics[scale=0.6]{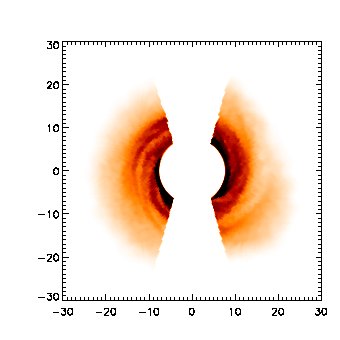}
\caption{Images of (a) raw HI data, and (b) a representative simulation (100R5 as denoted in CB09) that have symmetric wedges cut out (symmetric in the $\phi$ coordinate).  The box extends from -30 kpc to 30 kpc.}
\end{center}
\label{fig:symimages}
\end{figure}

\begin{figure}
\begin{center}
\includegraphics[scale=0.6]{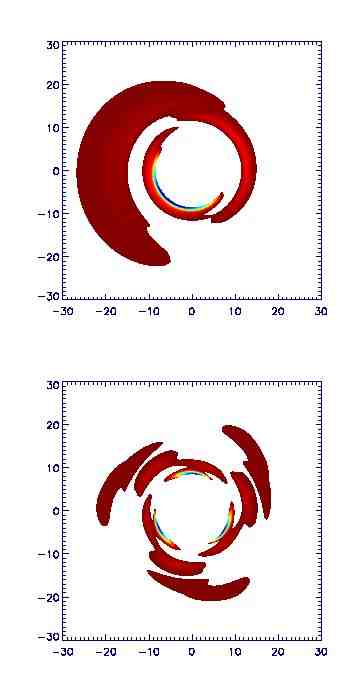}
\caption{Images of the odd modes in raw HI data (a) m=1, and (b) m=3.  The box extends from -30 kpc to 30 kpc.}
\end{center}
\label{fig:obsmodeimages}
\end{figure}

\begin{figure}
\begin{center}
\includegraphics[scale=0.6]{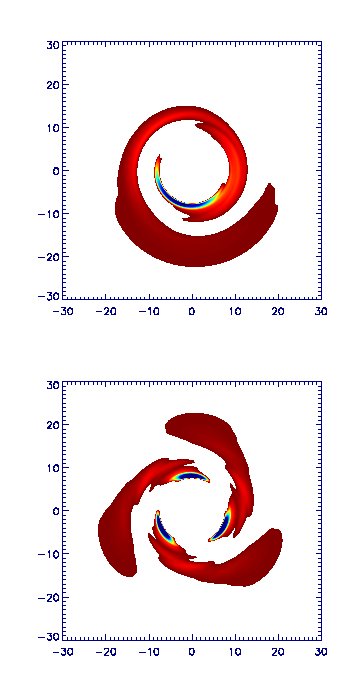}
\caption{Images of the odd modes in representative simulation 100R5 shown here at the best-fit time (a) m=1, and (b) m=3.  The box extends from -30 kpc to 30 kpc.}
\end{center}
\label{fig:simmodeimages}
\end{figure}

We focus our analysis of the phase variation on the $m=1$ and $m=3$ modes.  We do this firstly because, as we discuss in a longer forthcoming paper, the $m=1$ and $m=3$ modes are particularly affected by perturbers with mass ratios $\sim 1:100$ with $R_{\rm peri} \sim 5-10~\rm kpc$.  Secondly, the Fourier analysis, particularly the analysis of the phase of the modes, is affected by the wedges cut out of the HI data (see Levine, Blitz \& Heiles 2006b for the details of the size of the wedges).  The HI map cannot be reliably determined close to $l \sim 0,180$ because the velocities along the line of sight are too small with respect to random motions to establish reliable distances.  Therefore, the final map has wedges cut out of the data that are symmetric in galactic longitude but asymmetric in the galactocentric $\phi$ coordinate.  We have taken here the original HI map and rendered the wedges symmetric in $\phi$, as depicted in Figure 1 (a) for the raw HI data and in Figure 1(b) for a representative simulation (100R5).  It can be easily demonstrated that symmetric wedges affect the even modes primarily (as the wedges themselves contain even modes) using simple functions such as a constant function $f(r,\phi)=\rm constant$.  Thus, for an image with wedges cut out that are symmetric in $\phi$, the phase of the odd modes is less affected by the wedges than the even modes (as the wedges themselves contain even modes).  The 100R5 simulation was a case that fit the amplitudes of the Fourier modes of the data well; the images of 100R5 simulation shown in Figure 1b correspond to the best-fit time snapshot.  The images of the odd modes are shown in Figures 2 and Figure 3 for the raw HI data and simulation respectively.    


\begin{figure}
\begin{center}
\includegraphics[scale=0.5]{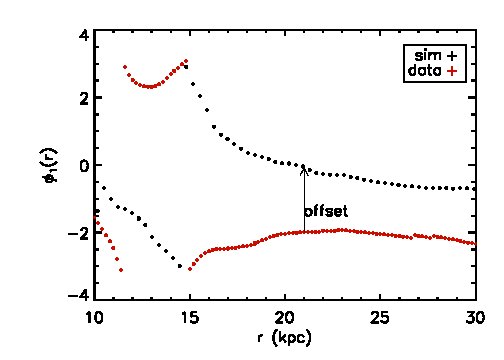}
\includegraphics[scale=0.5]{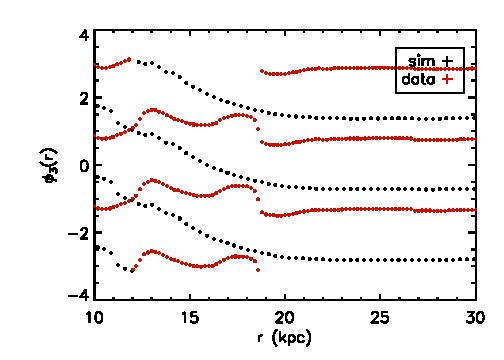}
\caption{Comparison of phase of (a) $m=1$ and (b) $m=3$ modes for 100R5 relative to raw HI data.  See text for description of how the azimuth of the perturber is determined from the $\it{relative~offset}$ between the phase of the simulations and data.  The phase of the simulations is shown here at the time that best-fits the Fourier amplitudes}
\end{center}
\label{fig:modeperturbedsims}
\end{figure}

Figure 4(a) and Figure 4(b) depict the phase variation of the $m=1$ and $m=3$ modes from a representative simulation (100R5), as compared to the phases of the raw HI data.  Of particular note is the steep variation of $\phi_{1}(r)$ in the inner regions ($r \la 15~\rm kpc$) where the dominant mode of spiral structure arises from the gas being forced by the stellar spiral arms inside of the Outer Lindblad Resonance (Chakrabarti et al. 2003; Chakrabarti 2009) for both the simulations and the data.  In contrast, $\phi_{1}(r)$ is relatively flat in the outer regions (for $r \ga 15~\rm kpc$), a signature of a tidal interaction (Shu 1984), where the wavelength of the disturbance increases outwards (similar to taking a bedsheet and flapping it).  In other words, tidal interactions produce open spiral planforms, while self-excited spiral structure is more tightly-wrapped.  The higher order modes have ``m'' branches at a given radius, while being similar to the $m=1$ mode in having a flat variation of the phase in the outer regions of the galaxy.  To determine the translation between the simulation frame and the observational frame, we simply determine the relative offset from the phases, i.e., $\delta\phi = \rm Median \left[\phi^{\rm sim}_{1}(r)-\phi^{\rm data}_{1}(r)\right]$.  The range of radii is restricted to the outer regions, between $19 \la  r \la 23 \rm kpc$; we have checked that taking the average or considering a range or radii somewhat different than this (as long as $r \ga 15~\rm kpc$ and $r \la 30~\rm kpc$ beyond which the data is not reliable) does not change the answer significantly.  Note that the phase comparison is done in the outskirts of the galaxy (for $r \ga R_{\rm OLR}$) to isolate the tidal response from the self-excited response, as the latter dominates in the inner regions.  For the $m=3$ mode, we have a priori three branches to select from to determine the offset; we select the branch of $m=3$ that is in magnitude closest to the $m=1$ branch for both the simulations and the data, in order to determine the offset.  In detail, we select the branch of $m=3$ by subtracting each branch of $\phi(r,m=3)$ from $\phi(r,m=1)$ and choosing the branch that gives the smallest median absolute difference (computed between $19~\rm kpc < r < 23~\rm kpc$).  This leads to, for instance, a median value of $\delta\phi=-1.9$ for $19~\rm kpc < r < 23~\rm kpc$ from the $m=1$ mode comparison, and $\delta\phi=-0.63$ from the $m=3$ mode comparison, when selecting the branch of $m=3$ closest to $m=1$ for both the simulations and data.  To determine the azimuth of the perturber, we employ the phase information from the images with symmetric wedges (as in Figure 1) as the information in the odd modes should be unaffected by symmetric wedges.  Figure 5 (a) and (b) display the phase of the odd modes in the simulations and data with symmetric wedges, which is slightly different from that shown in Figures 4, leading to $\delta\phi$ values derived from the $m=1$ and $m=3$ modes that are somewhat more comparable.

Finally, it is important to emphasize that the phase of the modes in the simulations is a function of time.  In the preceding figures (Figures 3-5), the phase is shown at the time that best fits the Fourier amplitudes of the data, which was determined earlier in CB09.  This best-fit time occurs approximately a dynamical time after pericentric approach.  The time variation of the phase also independently provides another handle on the time of encounter.  In fact, the phase of the modes in the simulations is flat in the outskirts at times close to the best-fit time, and assumes variable shape at times significantly preceding to or after the best-fit time. This is shown in the Appendix.


\begin{figure}
\begin{center}
\includegraphics[scale=0.5]{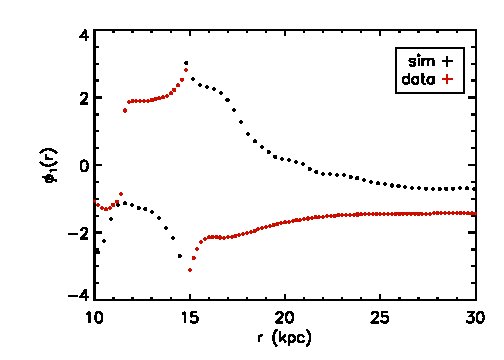}
\includegraphics[scale=0.5]{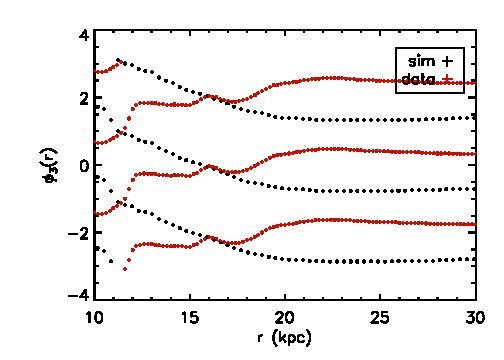}
\caption{Comparison of phase of (a) $m=1$ and (b) $m=3$ modes for 100R5 relative to raw HI data from images with wedges $\it{symmetric}$ in $\phi$.}
\label{fig:modesymmetrized}
\end{center}
\end{figure}

\subsection{Phase Variation:  Secular Vs. Tidal Response; Dependence on Initial Conditions}

\begin{figure}
\begin{center}
\includegraphics[scale=0.5]{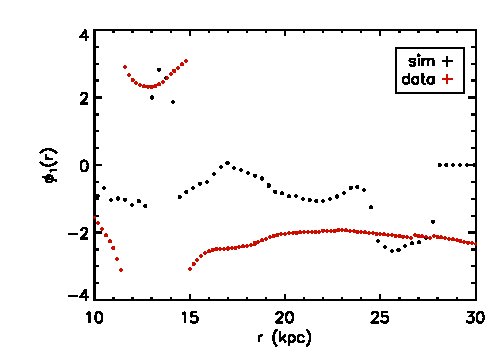}
\includegraphics[scale=0.5]{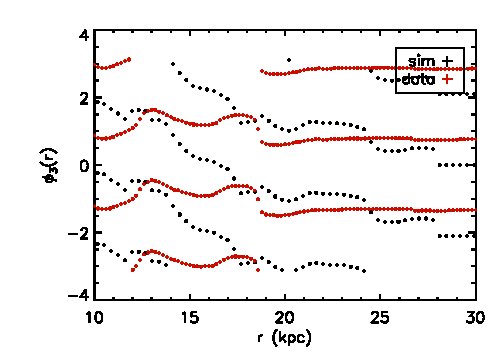}
\caption{Comparison of phase of (a) $m=1$ and (b) $m=3$ modes for an isolated galaxy simulation relative to raw HI data of the Milky Way.  The data (red) are the same as shown in Figure 4}
\label{fig:modeisosims}
\end{center}
\end{figure}

\begin{figure}
\begin{center}
\includegraphics[scale=0.5]{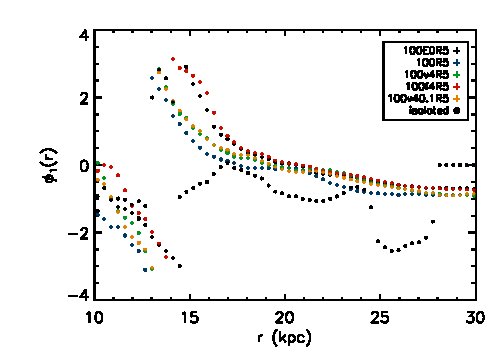}
\caption{Comparison of phase of $m=1$ mode for an isolated galaxy simulation relative to variants of 100R5}
\label{fig:modeisoallsims}
\end{center}
\end{figure}


\begin{figure}
\begin{center}
\includegraphics[scale=0.5]{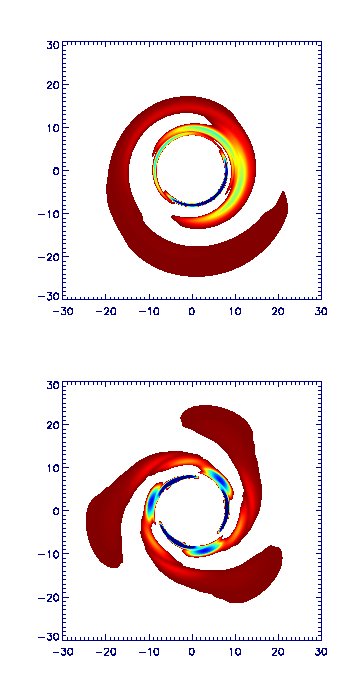}
\caption{Images of odd modes for the 100E0R5 simulation done with the isothermal equation of state, (a) $m=1$ and (b) $m=3$.  This should be compared to Figure 3 which shows the phase of odd modes for the 100R5 simulation which is performed with the SH03 multiphase ISM including energy injection from supernovae (see text for details).  Both of these simulations have very similar phase of the disturbances in the $\it{outskirts}$ although having different equation of state.}
\label{fig:modes100E0R5}
\end{center}
\end{figure}


\begin{figure}
\begin{center}
\includegraphics[scale=0.5]{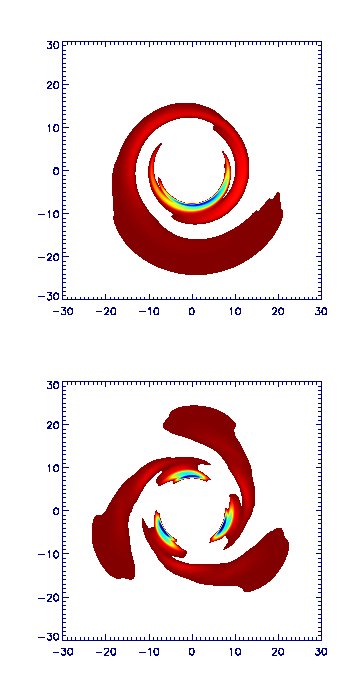}
\caption{Images of odd modes for the 100v4R5 simulation, (a) $m=1$ and (b) $m=3$}
\label{fig:modes100aR5}                                                                              
\end{center}
\end{figure}

We wish to underscore the visual signature of a tidal interaction that is present in Figures 4 and 5.  We do this by comparison of the phase of the odd modes of the raw HI data to that of an isolated galaxy simulation.  Figure 6(a) depicts the phase of the $m=1$ mode from an isolated galaxy simulation overplotted with the $m=1$ mode of the data, and Figure 6(b) shows the same for the $m=3$ mode.  As is clear, the $m=1$ mode of the isolated galaxy simulation does not the display the continuous flat profile as evinced by the 100R5 tidally interacting model; it discontinuously goes to zero in the outer regions.  The phase of the $m=3$ mode for the isolated galaxy simulation declines outwards, in contrast to that of the tidally interacting system shown in Figure 4(b), which is flat in the outer regions.  It is the case, as is expected, that the inner regions ($r \la 15~\rm kpc$) do display a gradient in $\phi_{1}(r)$ as the gas in the inner regions responds to the forcing by the stellar spiral arms.  It is unlikely that cosmologically grown asymmetrical dark matter haloes would reproduce both the amplitudes and phase of the Fourier modes in the absence of external perturbers, a point that we return to in \S 3.  

Secondly, we wish to highlight the relatively insensitive dependence of the phase of the modes on initial conditions for variants of the 100R5 simulations at the best-fit time to the Fourier amplitudes (where by variants, we mean variations in the initial conditions of the simulated Milky Way disk, with the parameters varied denoted in Table 1 of CB09).  Figure 7 depicts the phase variation of the $m=1$ mode of several of the 100R5 variants (for the nomenclature, see Table 1 of CB09) at the best-fit time over-plotted with the phase variation of the $m=1$ mode for an isolated galaxy simulation.  As is clear, the phase of the 100R5 variants at the best-fit time is quite similar in the outskirts of the galaxy ($R \ga 15~\rm kpc$) and distinct from that of the isolated galaxy simulation.  Figures 8 and 9 show the images of the odd modes and further illustrate the similarity of the phase of two of the 100R5 variants (100E0R5 and 100v4R5) at the best-fit time.  There are differences in the images of the modes for the isothermal 100E0R5 and the multiphase 100R5 simulation (shown in Figure 3) and the 100v4R5 simulation (shown in Figure 9) in the inner regions of the galaxy.  This is due to higher energy injection from supernovae in the multiphase 100R5 model thereby rendering it more stable, with the isothermal 100E0R5 model displaying a greater density contrast as it is more susceptible to fragmentation without the energy injection from supernovae.   All of the images are however remarkably similar in the outskirts.  The reason this is so is that in the Springel \& Hernquist (2003) multiphase ISM model, the energy injection is proportional to the star formation rate.  In the outskirts of the galaxy, there is very little star formation, and as such little difference arises due to variations in the equation of state of the gas.  Due to the relative lack of star formation in the outskirts of galaxies (i.e., relative to the inner regions; though see Thilker et al. 2005 for observations of UV emission that indicate a low level of star formation in the outskirts of galaxies), the study of gas dynamics in the outskirts of galaxies is comparatively simpler as it is less dependent on processes such as star formation and feedback that are difficult to model in detail and at present generally prescribed in a sub-grid manner. 

\subsection{Radial Velocity Constraint: $v_{\pi}(R)$}


\begin{figure}
\begin{center}
\includegraphics[scale=0.4]{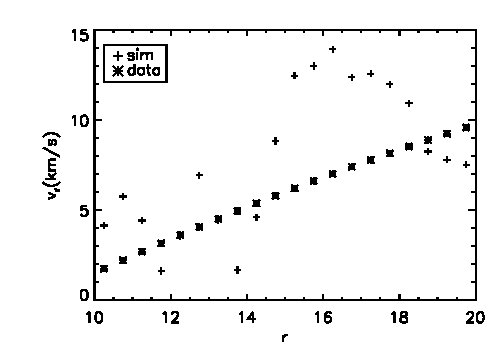}
\includegraphics[scale=0.4]{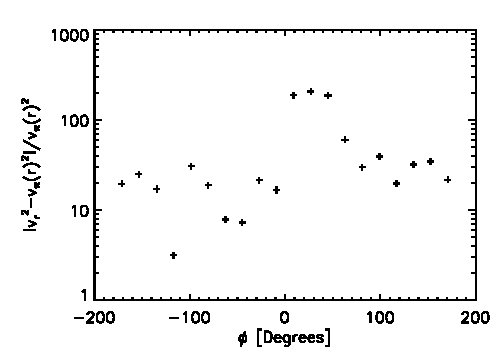}
\caption{(a) comparison to $v_{\pi}(R)$ for best-fit $\phi$ value, (b) distribution of $\left[v_{r}^{2}-v_{\pi}(R)^{2}\right]/v_{\pi}(R)^{2}$ as a function of the simulation $\phi$ coordinate at best-fit time}
\end{center}
\label{fig:velocity}
\end{figure}

Levine, Blitz \& Helies (2006b) define a quantity $v_{\pi}(R)$ that represents the radial velocity at $l=180$, and allows one to correct for the asymmetry between the surface densities at Galactic longitudes on either side of $l=0$ and $l=180$ (Henderson et al. 1982).  In essence, the quantity $v_{\pi}(R)$ represents the magnitude of the elliptical corrections to circular motions.  Assumption of pure circular motions produces a surface density map that is discontinuous on either side of $l=0$ and $l=180$.  Levine et al. (2006b) derive the functional form of $v_{\pi}(R)$ by requiring continuity of the surface density in these regions.  One may assume that this asymmetry in the radial velocity field arises from a perturber forcing the gas in the outskirts.  If so, $v_{\pi}(R)$ potentially gives us another estimate of the azimuth of the perturber, in addition to that already given by the relative offset in the phases of the surface density maps of the simulations and the data.

In Figure 10, we compare the radial velocity for a representative simulation (100R5) at the time snapshot that gives the best-fit to the amplitudes of the Fourier modes of the raw HI data.  Specifically, Figure 10 (a) shows the comparison of $v_{\pi}(R)$ (in asterisks) and $v^{\rm sim}_{r}$ (crosses) at the best-fit azimuth.  The general behavior of the two curves is in agreement, although the simulation curve overestimates the radial velocity by $\sim 2$ in the outer regions, relative to $v_{\pi}(R)$ in the outer regions.  It is important to note that the functional form of $v_{\pi}(R)$ as given in Levine, Blitz \& Heiles (2006b) is itself an average and has some intrinsic scatter.  As such, disagreement at the $\sim 2$ level may not be significant.  It is difficult to refine this comparison further due to lack of the full velocity field.  We address this problem in a future paper for THINGS galaxies where we compare both the HI and velocity maps with the analogous simulation quantities.  

There is a reasonable level of agreement between $v_{r}^{\rm sim}$ and $v_{\pi}(R)$ at the best-fit angle. When translated to sun-centered coordinates this best-fit angle agrees with the azimuth calculated from the phase offset to within $\sim$ 15 degrees for the variants of the 100R5 simulations (variations in initial conditions, orbits etc represent the variants in 100R5 as discussed in CB09).  Figure 10(b) shows that the square of $v_{r}^{\rm sim}$ deviates from the square of $v_{\pi}(R)$ significantly at all angles except at the best-fit angle.   This suggests that the asymmetry in the radial velocity field can be effectively analyzed in this way to find the best-fit angle for the perturber, which corresponds to the $\phi$ value at which $\left[|v_{r}^{2}-v_{\pi}(R)^{2}|\right]/v_{\pi}(R)^{2}$ is a minimum.  We also note that the simulations of isolated galaxies differ from tidally interacting galaxies not only in the phase variation in the outskirts of the galaxy (with tidally interacting galaxies displaying open spirals or a radially flat variation of the phase), but also in that the isolated galaxy simulations do not show as good an agreement with $v_{\pi}(R)$.  Specifically, if we take the quantity $\mathcal{M}\equiv\frac{|v_{r}^{2}-v_{\pi}(R)^{2}|}{v_{\pi}(R)^{2}}$ as our metric of comparison, then variants of 100R5 have a minimum in $\mathcal{M}$ of order unity, while isolated galaxies have minima in $\mathcal{M} \ga 10$, indicating that the 100R5 tidally interacting systems give a better fit overall to the observed asymmetry in the radial velocity field.  Although the observations of the Milky Way do not provide the full velocity field, observations of galaxies that display a clear asymmetry in the velocity field in the outer regions are likely to be tidally interacting systems (Dobbs et al. 2010).


\begin{figure}
\begin{center}
\includegraphics[scale=0.35]{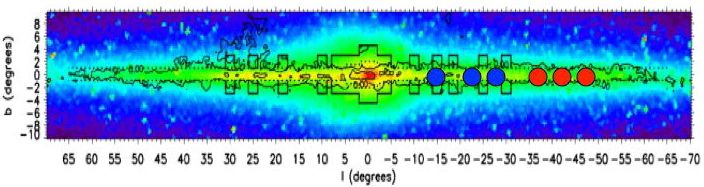}
\caption{Distribution of predicted ``l'' values of perturber as derived from comparison to $m=1$ mode (in red) and from comparison to $m=3$ mode (in blue).}
\end{center}
\label{fig:lvalues}
\end{figure}

We now describe the coordinate transformation performed to calculate the azimuth of the perturber in sun-centered coordinates.  We take the sun-Galactic center distance $R_{0}=8.5~\rm kpc$, and the radial location of the perturber at the best-fit time as determined by CB09 to be $R \sim 80~\rm kpc$.  From the analysis of CB09, we also know the perturber azimuth in simulation coordinates in addition to the current radial location of the perturber.  We denote the azimuth of the sub-halo in simulation coordinates as $\phi_{\rm sat}^{\rm sim,avg}$ which is an average value over the 100R5 variants (with a range of a degree) equal to 217 degrees in center-of-mass coordinates.  The angle of the perturber in galactocentric coordinates as determined by the relative offset in the phase is given by: $\phi_{\rm perturber}=\phi_{\rm sat}^{\rm sim,avg}-\phi_{m}^{\rm sim}+\phi_{m}^{\rm data}$, where the $\phi_{m}'s$ are the phases of modes ``m'' in the simulations and data.  Note that the $\phi_{m}'s$ are functions of $r$, while the quantity $\delta\phi = \rm Median\left[\phi_{m}^{\rm sim}(r) - \phi_{m}^{\rm data}(r)\right]$ is a number.  With these quantities and simple geometry, we then find that the longitude ``l'' is given by:
\begin{eqnarray}
l=asin\left[\frac{R}{r'}sin(90-\phi_{\rm perturber})\right] \\
r'=\left[R_{0}^{2}+R^{2}-2RR_{0}cos(90-\phi_{\rm perturber})\right]^{1/2}   \; ,
\end{eqnarray}
which gives $-50 \la l \la -10$ for the best-fit cases.  The derived azimuth values for the perturber are shown in Figure 11 overlaid on a map of the Milky Way.

\section{Discussion}

Our methodology to characterize dark sub-halos here is a simple one -- in essence, we found in CB09 that the mass of the perturber was given by the amplitude of the Fourier modes, and disturbances were excited on scales of order $R_{0}(M_{s})$ (the minimum distance at which a satellite of mass $M_{s}$ starts to affect a parcel of the primary galaxy so as to induce radial motions of order the sound speed).  This leads to therefore disturbances being produced on larger scales for more massive satellites (see Figure 3 in CB09) as their $R_{0}$ is larger -- producing a visually different response for two satellites of differing mass that exert the same tidal force at their respective pericentric approach distances.  The nonlinear response of gas in the primary galaxy along the course of the orbit of the perturber (from $R_{0}(M_{s})$ to $R_{\rm peri}$) allow us to break the degeneracy between the mass of the perturber and distance of closest approach.  If $R_{\rm peri}/R_{\rm s} \gg 1$ (where $R_{\rm s}$ is the scale length of the galaxy) then the condition stated in CB09 cannot be realized, i.e., the time it takes the satellite to traverse from $R_{0}$ to $R_{\rm peri}$ becomes infinitesimally small as $R_{0}$ approaches $R_{\rm peri}$ in this limit (for non-major mergers).  Similarly, for smaller and smaller masses, $R_{0}(M_{s})$ approaches $R_{\rm peri}$ and the degeneracy cannot be broken.  However, it is the regime where the pericentric distance is of order or a few times greater than the scale length wherein a significant tidal force is exerted -- it is in this regime that CB09 found the $M,R^{3}$ degeneracy can be broken.  

We have utilized three separate and $\it{independent}$ constraints to characterize dark perturbers in this paper:  1) the radial variation of the Fourier modes (to determine the mass and pericentric distance -- these are results we utilize from CB09), 2) the phase of the Fourier modes (to determine the azimuth of the perturber from the relative offset in phase between the simulations and data), and 3) the asymmetry in the radial velocity field, which yields another measure of the azimuth of the perturber in addition to that derived from the relative offset of the phase.  We also find that interactions of $1:100$ satellites with $R_{\rm peri} \sim 5-10~\rm kpc$ preserve the observed constraint of the thin stellar disk of the Milky Way, in agreement with prior studies (Purcell et al. 2009; Kazantzidis et al. 2008).  The physical reason for this is that the sub-halos that we consider are not point masses -- they have some physical extent and circular velocity, which mediates the amount of energy transfer to the stellar disk.  Specifically, $\sim 1:100$ sub-halos have a circular velocity (at $\sim R_{\rm vir}$) that is comparable to the velocity dispersion of the galactic stellar disk.  For extended bodies the maximum velocity that can be imparted is the escape velocity -- as such, a kick from such sub-halos can at most produce an order unity effect as the velocity imparted will be stabilized by a comparable stabilizing force due to the intrinsic velocity dispersion of the stellar disk.  

Finally, it is not unreasonable to posit that these disturbances arise from a satellite close to the galactic plane -- dust obscuration close to the plane of the galaxy, to within $\sim$ 5 degrees of the galactic plane, renders optical identification of dwarf galaxies difficult (using dust extinction from Schlegel, Finkbeiner \& Davis 1998).  Even though a $10^{10}M_{\odot}$ satellite is almost comparable in mass to the largest known satellites of the Milky Way, it is important to emphasize that this mass estimate refers to the total mass, and not the stellar mass.  The mapping between the the dark matter halo mass and the stellar mass is not a linear relation and is influenced by a number of processes, in particular the formation redshift.  Stewart (2009) find that the stellar to dark matter mass at a fixed halo mass decreases for halos formed at higher redshift.  Thus, it is conceivable that a $10^{10}M_{\odot}$ halo formed at $z \sim 1$ will have a lower light-to-mass ratio than one formed more recently, and be more difficult to detect especially if it is close to the galactic plane. 

We comment briefly on the possibility of the Sagittarius dwarf galaxy (Sgr) as being the culprit that produced these disturbances.  We noted in CB09 that the LMC is too far away to raise sufficient tides on the galactic disk, and that it is more difficult to rule out Sgr as the dominant architect of these disturbances.  Cited masses for Sgr range from $\sim 10^{9}~M_{\odot}$ to $\sim 10^{11}~M_{\odot}$ (Jiang \& Binney 2000) although more recent work (Niederste-Ostholt et al. 2010) that analyzes the stellar debris from SDSS and 2MASS cites a mass of $\sim 10^{10}~M_{\odot}$. If the more recent numbers are considered, then the crucial quantity is the last pericentric approach, i.e., not the current one at 24 kpc from the Sun (Ibata et al. 1997) but the previous approach as the disturbances manifest about a dynamical time after pericentric approach.  As noted also in CB09, earlier calculations (e.g. Johnston et al. 1999) find pericentric approach at the prior passage to be $\sim 40~\rm kpc$.  If it is possible to have a closer pericentric approach at $\sim 10~\rm kpc$, then Sgr becomes a viable candidate on the basis of comparison to the Fourier amplitudes.  Our analysis here based on the phase of the disturbances find azimuth values to be different from the present location of Sgr.

It is altogether possible that a more detailed model, including a non-spherical halo as found in cosmological simulations (Allgood et al. 2006; Kuhlen et al. 2007) and the inclusion of magnetic instabilities, may well yield a better fit to one of these constraints.  It is unlikely however, that all three of these constraints can be fit significantly better.  We do consider in future papers the effects of deviations from spherical symmetry in the dark matter halo on the generation of these disturbances, as well as the effect of multiple perturbers, and magnetic instabilities.  One may expect that the phase variation of the modes when multiple perturbers are interacting with the galactic disk will be less coherent than that produced by a single perturber.  The relatively coherent phase of the raw HI data does not immediately point to the need for multiple perturbers.  However, in principle this is a topic that deserves close attention due to the prevalence of multiple perturbers in cosmological simulations, which may produce a strong signal if the distribution in time, mass and radius are such to excite tidal interactions.  It bears mention that if the distribution of sub-halos follows a power-law as found in cosmological simulations (Diemand et al. 2007), the tidal heating and therefore the resultant Fourier amplitudes of the gas disturbances will be dominated by the most massive perturbers.  If multiple sub-halos tidally interact with the galactic disk $\it{within}$ a dynamical time, their effect will be sub-linear and will saturate due to heating of the galactic disk from successive encounters.  If multiple sub-halos interact with the disk only infrequently (every few dynamical times) their effect will be washed out in the gaseous component, as the cold gas is dissipative, not retaining much long-term memory of past dynamical encounters.  Thus, while the most probable scenario is one where a single perturber dominates the signal, the effect of multiple sub-halos will likely contribute a low level to the overall Fourier amplitudes, the magnitude of which will be useful to determine.

We discuss briefly the consideration of a triaxial halo as it has received considerable attention in the literature.  Generally, dark matter haloes are observationally inferred to be rounder than what is predicted by collisionless cosmological CDM simulations, from gravitational lensing studies (Mandelbaum et al. 2006) and X-ray isophotal studies (Buote et al. 2002).  This may be due to the condensation of baryonic components leading to rounder halos (Kazantzidis et al. 2004a; Debattista et al. 2008), or possibly due to uncertainties in the observational determination of the triaxiality of halos as the observational inferences are based on the projected mass distribution.  Therefore, while the investigation and determination of discriminants between secular effects, due to a triaxial dark matter halo, and those from external perturbers is a topic worthy of serious investigation, it is not immediately clear that the prevalence of non-spherical halos found in collisionless cosmological simulations is reflected in reality.  Nonetheless, a comprehensive understanding of the differences between secular effects with the inclusion of non-spherical halos and external perturbers would be very useful.  However, it is difficult to produce stable initial conditions for multi-component disks with non-spherical halos, especially with the inclusion of gas.  Such a study is beyond the scope of the present paper, and the presence of this possibility may be considered a caveat to the model presented here. 

\section{Conclusion}

$\bullet$  We continue our development of the ``tidal analysis'' method of CB09 to characterize dark perturbers from observed disturbances in gas disks.  We examine here the phase of the observed disturbances of the Milky Way HI disk, in addition to the Fourier amplitudes that were analyzed by CB09.  As shown in CB09, the Fourier amplitudes yield the mass and pericentric distance of the perturber.  We examine here the phase variation of the low order modes in both the simulations and the data with specific attention to the odd modes to find that the phase of the modes in the outer regions (for $r \ga 15~\rm kpc$) is relatively flat as a function of radius, i.e., qualitatively of the form of a tidal interaction.  This is true for the variants of the 100R5 simulations at the best-fit time (the time that best fits the Fourier amplitudes of the data as determined in CB09).  The phase variation for both the simulations and the data show a sharp gradient in the inner regions (as expected when gas is forced by the stellar spiral arms).  To translate between the observed frame and the simulation frame, we find the relative offset between the phases of the modes, and use this offset to calculate the azimuthal location of the perturber, in both galactocentric and sun-centered coordinates.  We find that the perturber responsible for the disturbances in the HI disk of the Milky Way is at a longitude $-50 \la l \la -10$, assuming that that the angles calculated from the modes are independent.

$\bullet$ The phase variation of the simulated galaxies at the best-fit time is largely insensitive to the initial conditions.  This lack of sensitivity to the initial conditions (specifically, the equation of state) for the phase of the modes in the outskirts of the simulated galaxy arises from the relative lack of star formation (i.e., relative to the inner regions) which serves to set the effective equation of state and energy injection from supernovae.  Therefore, even though global simulations of galaxies presently treat star formation in a simple, prescriptive manner and the equation of state of the gas is an unknown parameter, one can reach conclusions about the evolution of gas on the outskirts of galaxies that are largely independent of processes such as star formation and feedback that are difficult to model in detail. 

$\bullet$  We have compared the radial velocity field in the simulations to find not only the angle at which $v_{r}$ gives the best-fit to the asymmetry in radial velocity ($v_{\pi}(R)$) as observed by Levine et al. (2006), but also the distribution of chi-squared values as a function of angle.  The angle at which $v_{r}$ most closely matches $v_{\pi}(R)$ is comparable to that derived from the phase comparison, to within $\sim$ 15 degrees.  Moreover, the distribution of chi-squared values of $v_{r}$ in the simulation as a function of angle increases sharply when moving away from the best-fit angle.  

$\bullet$ We emphasize that our calculation of the azimuth of the perturber derives from three separate and $\it{independent}$ constraints (the Fourier amplitudes, the phase and the asymmetry in the radial velocity) that are fit reasonably well within the context of our adopted model of an external perturber.

\bigskip
\section*{Acknowledgements}

SC thanks Greg Laughlin for inspiring discussions.  We also thank Chris McKee, Mark Krumholz, James Bullock, Chris Purcell, Eliot Quataert, Chung-Pei Ma, Fabio Governato and Lucio Mayer for helpful discussions.  We thank Phil Chang for a careful reading of the manuscript and constructive comments, and especially CK Chan for insightful discussions on Fourier analysis.  SC is supported by an UC President's Fellowship.  



\appendix
\section{Time Variation of Simulation Phase}

As noted earlier in \S 2, the time variation of the phase independently provides a handle on the time of encounter, just as the time variation of the Fourier amplitudes allows one to constrain the time of encounter.  In this paper, we performed our comparison of the phase of the modes at the best-fit time, i.e., the time at which the Fourier amplitudes of the data best-fit those of the simulations.  It bears mention that it is at (or close to) this time that the phase of the modes is in shape most similar to that of the HI data.  Thus, one could in principle  determine the time of encounter equivalently by reference to the shape of the phase of the modes in comparison to that of the data.  Within the context of our model, these two independent constraints yield the same time of encounter.  

Figure A1 depicts the gas surface density images of the 100R5 simulation as a function of time with the centers of the primary galaxy and perturber marked by crosses. Pericentric approach occurs at $t=0.29~\rm Gyr$ when the perturber is 5 kpc from the primary galaxy, and the best-fit to the Fourier modes occurs at $t=0.55~\rm Gyr$ by which point the perturber is 80 kpc from the center of the primary galaxy.  Figure A2 shows the phase of the $m=1$ mode at $t=0.025 ~\rm Gyr$ (this is prior to pericentric approach), at $t=0.55~\rm Gyr$ (at the time that best fits the Fourier amplitudes of the data as determined in CB09), and at $t=1~\rm Gyr$ (a few dynamical times after pericentric approach).  It is clear that well before the best-fit time to the Fourier amplitudes (at $t=0.025 ~\rm Gyr$) the phase is more comparable to that of an isolated galaxy as shown in Figure 6, while at $t=0.55~\rm Gyr$ the phase is nearly flat in the outskirts as is the case for the HI data.  Well after the best-fit time (at $t=1~\rm Gyr$) the spirals wind up on the outskirts, turning the nearly flat phase on the outskirts of the galaxy at the best-fit time to one with a more sharp gradient.  Thus, cold gas, due to its dissipative nature, is a particularly useful tracer in allowing for the identification of the time of encounter from either the Fourier amplitudes or the phase.  Before pericentric approach the phase does not display the nearly flat shape that is characteristic of the best-fit time, while well afterwards the phase declines outwards as the spirals wind up.  The nearly flat variation of the phase in the simulations -- which agrees closely in shape to the phase of the HI data -- occurs at a time that is coincident with the best-fit to the Fourier amplitudes of the HI data.  It is therefore reassuring that our model yields the same time of encounter from two independent constraints.

\begin{figure}
\begin{center}
\includegraphics[scale=0.4]{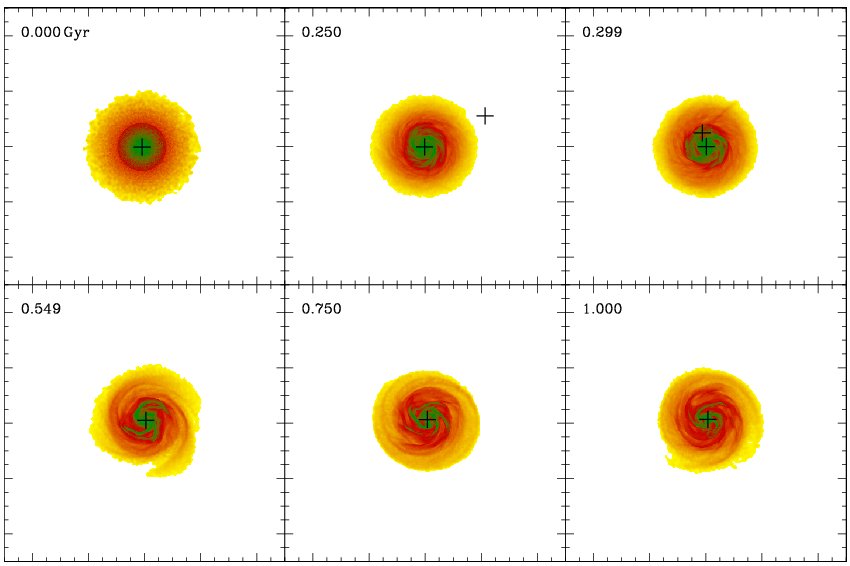}
\caption{Gas surface density images of 100R5 as a function of time, with crosses marking the centers of the primary galaxy and the perturber.  Length of the box is 50 kpc.  Note that pericentric approach occurs at $t=0.29~\rm Gyr$ and the best-fit to the Fourier amplitudes occurs at $t=0.55~\rm Gyr$.}
\end{center}
\end{figure}

\begin{figure}
\begin{center}
\includegraphics[scale=0.5]{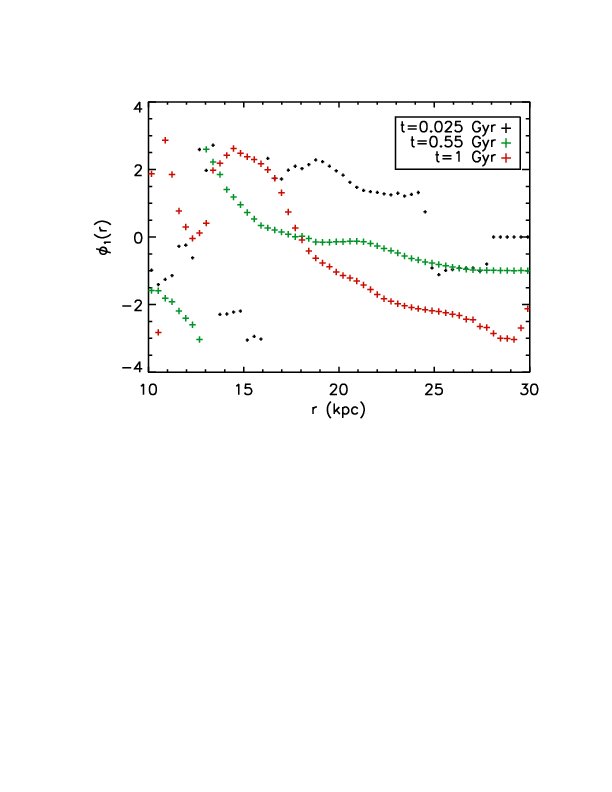}
\caption{Phase of the $m=1$ mode in the 100R5 simulation shown at three times: 1) at $t=0.025~\rm Gyr$, prior to pericentric approach when the perturber has not yet reached pericentric approach; 2) at $t=0.55~\rm Gyr$ - this is about a dynamical time after pericentric approach when the Fourier amplitudes of the simulation closely match that of the observed data, and 3) $t=1~\rm Gyr$ -- this is a few dynamical times after pericentric approach.}
\end{center}
\end{figure}


\begin{thebibliography}{99}

\bibitem[]{}Allgood, B., et al., 2006, MNRAS, 367, 1781A
\bibitem[]{}Belokurov, V., et al., 2006, ApJ, 647L, 111B
\bibitem[]{}Block,D.L., et al. 2009, ApJ, 694:115-129
\bibitem[]{}Buote,D. et al., 2002, ApJ, 577, 183B
\bibitem[]{}Chakrabarti, S. \& Blitz, L., 2009, MNRAS Letters, 399L, 118C [CB09]
\bibitem[]{}Chakrabarti,S.,Laughlin,G. \& Shu, F.H., 2003, ApJ, 596, 220-239
\bibitem[]{}Chakrabarti,S.,2009, submitted to MNRAS, arXiv: 0812.0821C
\bibitem[]{}Considere, S. \& Athanassoula, E., 1982, A\&A, 111, 28C
\bibitem[]{}Dalal,N. \& Kochanek, C., 2002, ApJ, 572:25-33
\bibitem[]{}Debattista, V.P., et al. 2008, ApJ, 681, 1076
\bibitem[]{}Diemand,J., Kuhlen, M. \& Madau, P., 2007, ApJ, 667, 859
\bibitem[]{}Garcia-Gomez, C. et al. 2004, A\&A, 421, 595
\bibitem[]{}Governato,F., et al., 2010, Nature, 463, 203G
\bibitem[]{}Henderson, A.P. et al. 1982, ApJ, 263, 116H
\bibitem[]{}Ibata,R. et al., 1994, Nature, 370, p. 194
\bibitem[]{}Jiang, I. \& Binney, J., 2000, MNRAS, 314, 468J
\bibitem[]{}Kazantzidis, S., Kravtsov, A. et al., 2004a, ApJ, 611L, 73K
\bibitem[]{}Kazantzidis,S.,Bullock,J.S. et al. 2008, ApJ, 688, 254-276
\bibitem[]{}Klypin,A.,Kravtsov,A.V.et al. 1999, ApJ, 522: 82-92
\bibitem[]{}Kravtsov,A., et al., 2004, ApJ, 609, 35K
\bibitem[]{}Kuhlen,M., Diemand, J. \& Madau, P., 2007, ApJ, 671, 1135K
\bibitem[]{}Levine,E.S.,Blitz,L.\& Heiles,C.,2006a, Science, 312, 1773L 
\bibitem[]{}Levine,E.S.,Blitz,L.\& Heiles,C.,2006b, ApJ, 643, 881L
\bibitem[]{}Mandelbaum,R. et al., 2006, MNRAS, 370, 1008
\bibitem[]{}Moore,B.,Ghigna,S. et al. 1999, ApJ, 524, L19-L22
\bibitem[]{}Purcell,C., Kazantzidis, S. \& Bullock, J.S., 2009, ApJ, 694L, 98P
\bibitem[]{}Shu,F.H.,1984,Planetary Rings(A85-34401 15-88),University of Arizona Press, Tucson, AZ, p. 513-561
\bibitem[]{}Springel,V.,2005,MNRAS,364,1105S
\bibitem[]{}Springel,V. \& Hernquist, L., 2003, MNRAS, 339, 289S
\bibitem[]{}Stewart,K.R.,2009,ASPC,419,243S
\bibitem[]{}Thilker, D.A., et al., 2005, ApJ, 619L, 67T
\bibitem[]{}Willman, B. et al., 2005, ApJ, 626L, 85W
\end{thebibliography}
\end{document}